# Large payload quantum steganography based on cavity quantum electrodynamics

Tian-Yu Ye, Li-Zhen Jiang

College of Information & Electronic Engineering, Zhejiang Gongshang University, Hangzhou 310018, P.R.China

A large payload quantum steganography protocol based on cavity quantum electrodynamics (QED) is presented in the paper, which effectively uses the evolution law of atom in cavity QED. The protocol builds up hidden channel to transmit secret messages using entanglement swapping between one GHZ state and one Bell state in cavity QED together with the Hadamard operation. The quantum steganography protocol is insensitive to cavity decay and thermal field. The capacity, imperceptibility and security against eavesdropping are analyzed in detail in the protocol. It turns out that the protocol not only has good imperceptibility but also possesses good security against eavesdropping. In addition, its capacity of hidden channel achieves five bits, larger than most of those previous quantum steganography protocols.

**Key words:** quantum steganography, hidden capacity, entanglement swapping, Hadamard operation, cavity QED

## Introduction

Steganography is a technique that hides secret message into innocent-looking cover data to make stego data almost the same as cover data. In essence, steganography hides the existence of secret message to escape from the suspicion of an eavesdropper. Cryptography is the art of scrambling secret message. In essence, cryptography hides the content of secret message. However, secret message scrambled by cryptography is so unreadable that it easily arouses the suspicion of an eavesdropper.

Quantum secure communication has been one of the most ingenious applications of quantum mechanics, which is unconditionally secure. It mainly includes quantum key distribution (QKD)[1-2], quantum secure direct communication (QSDC) [3-9],quantum secret sharing (QSS) [10-14]and so on. The aim of QKD is to establish an unconditionally secure key between two remote legitimate parties through the transmission of quantum signal. In 1984, Bennett and Brassard[1] proposed the first QKD protocol (BB84 protocol) using single particle. In 2002, Long et al.[2]proposed an efficient two-step QKD protocol. QSDC's object is to transmit the secret message directly between two remote legitimate parties via quantum channel without establishing a key in advance. In 2002, Beige et al.[3]proposed the first QSDC protocol. Then, Bostrom and Felbinger[4]proposed the famous Ping-Pong protocol (BF protocol) in 2002. In 2004, Cai and Li[5] proposed an improved Ping-Pong protocol (IBF) to double the capacity of Ping-Pong protocol by introducing two additional unitary operations. In 2008, Chen et al.[6] proposed a novel controlled QSDC protocol with quantum encryption using a partially entangled GHZ state. In 2008, Chen et al.[7] proposed a novel three-party controlled QSDC protocol based on W state. In 2011, Gu et al.[8] proposed a two-step QSDC protocol with hyperentanglement in both the spatial-mode and the polarization degrees of freedom of photon pairs. In 2012, Huang et al.[9] proposed two robust channel-encrypting QSDC protocols over different collective-noise channels. QSS is generalization of classical secret sharing to a quantum scenario, where only all parties cooperate together can they obtain the secret message sent by the initial sender through the transmission of quantum signal. In 1999, Hillery et al.[10] proposed the first QSS protocol based on Greenberger- Horne-Zeilinger (GHZ).In 2003, Guo et al.[11] proposed a QSS protocol only using product states. In 2010, Yang et al.[12] put forward a novel three-party QSS protocol of secure direct communication based on $\chi$-type entangled states. In 2012, Zhu et al.[13] made an improvement to the QSS protocol in Ref. [12] to prevent Bob form legally obtaining secret messages without being detected. In 2012, Chen et al.[14] put forward a three-party QSS protocol via the entangled GHZ state using the single-particle quantum state to encode the information. In recent years, based on the efforts of researchers, quantum secure communication, including QKD, QSDC, QSS and so on, has been greatly developed.

In the meanwhile, quantum data hiding (QDH)[15-16] is also presented, which derives from QSS. The feature of QDH is that some parties are unable to extract secret message only by using local operations together with classical communication. Quantum steganography[17-22], regarded as a new branch of quantum secure communication in recent years, is the generalization of classical steganography to a quantum scenario. It always builds up a hidden channel within normal quantum channel to transmit secret message, and can be applied in covert communication, quantum identity authentication, and so on. In 2002, Gea-Banacloche[17] encoded the arbitrary quantum data using a quantum error-correcting code(QECC) and hided secret message as errors. In 2004, Worley III[18] presented a fuzzy quantum watermarking scheme based on the relative frequency of error in observing qubits in a dissimilar basis from the one where they were written. In 2007, Martin[19] presented a quantum steganography protocol based on the BB84 protocol[1]. In 2010, Liao *et al.*[20] proposed a multi-party quantum steganography protocol based on Guo *et al.*'s QSS protocol[11]. In 2010, Qu *et al.*[21] presented a new quantum steganography protocol with large payload based on the improved ping-pong protocol (IBF)[5]. In 2011, Qu *et al.*[22] presented a new quantum steganography protocol with large payload based on entanglement swapping of $\chi$-type entangled states.

However, most of the above quantum steganography protocols[17-21] have the drawback of small hidden capacity. The hidden capacity of Ref.[17] is only one bit or one qubit. The hidden capacity of Ref.[18], Ref.[19] and Ref.[20] is also only one bit, respectively. We think that the capacity of hidden channel in both Ref.[17], Ref.[18], Ref.[19] and Ref.[20] is too small to efficient covert communication. Although the capacity of hidden channel in Ref.[21] can achieve four bit per round covert communication, we still think that it should be further improved.

In this paper, we aim to enhance the capacity of hidden channel, since it is a key point for an efficient quantum steganography protocol. A large payload quantum steganography protocol based on cavity quantum electrodynamics (QED) is proposed in the paper, which effectively uses the evolution law of atom in cavity QED. The protocol builds up hidden channel to transmit secret messages using entanglement swapping between one GHZ state and one Bell state in cavity QED together with the Hadamard operation. Its capacity of hidden channel achieves five bits, larger than that of Ref.[17], Ref.[18], Ref.[19], Ref.[20] and Ref.[21].

## 2 Encoding Scheme

As two-atom maximally entangled states, Bell states form a complete orthogonal basis for four-dimensional Hilbert space. Four Bell states are $|\phi^{\pm}\rangle = (1/\sqrt{2})(|gg\rangle \pm |ee\rangle)$ and $|\psi^{\pm}\rangle = (1/\sqrt{2})(|ge\rangle \pm |eg\rangle)$, respectively. As three-atom maximally entangled states, GHZ states form a complete orthogonal basis for eight-dimensional Hilbert space. Eight GHZ states are $|S^{\pm}\rangle = (1/\sqrt{2})(|gee\rangle \pm |egg\rangle)$, $|P^{\pm}\rangle = (1/\sqrt{2})(|ggg\rangle \pm |eee\rangle)$, $|Q^{\pm}\rangle = (1/\sqrt{2})(|gge\rangle \pm |eeg\rangle)$ and $|R^{\pm}\rangle = (1/\sqrt{2})(|geg\rangle \pm |ege\rangle)$, respectively.

Assume that the atoms $A$, $B$, $C$ are in one of the four GHZ states including $|S^{\pm}\rangle_{ABC} = (1/\sqrt{2})(|gee\rangle \pm |egg\rangle)$ and $|P^{\pm}\rangle_{ABC} = (1/\sqrt{2})(|ggg\rangle \pm |eee\rangle)$, while the atoms $D$, $E$ are in one of the above four Bell states. The atoms $A$, $D$ belong to Alice, and the atoms $B$, $C$, $E$ belong to Bob. Alice aims to covertly send secret messages to Bob by quantum channel. $U_0 = I = |g\rangle\langle g| + |e\rangle\langle e|$, $U_1 = \sigma_X = |g\rangle\langle e| + |e\rangle\langle g|$, $U_2 = i\sigma_Y = |g\rangle\langle e| - |e\rangle\langle g|$ and $U_3 = \sigma_Z = |g\rangle\langle g| - |e\rangle\langle e|$ are four unitary operations, respectively. Each unitary operation represents two bits information as formula (1). $U_0 \to 00$, $U_1 \to 01$, $U_2 \to 10$, $U_3 \to 11$. (1)

It is apparent that after performed one of the four unitary operations, the above four Bell states can be transformed into each other. It is also straightforward to the above four GHZ states.

Without loss of generality, assume that the initial state of the atoms $A$, $B$, $C$ is $|S^{-}\rangle_{ABC}$, and the initial state of the atoms $D$, $E$ is $|\psi^{-}\rangle_{DE}$. Two identical single-mode cavities are considered here. The atoms $A$, $D$ are sent to Bob by Alice through quantum channel at first. Afterward, the atoms $A$, $D$ are simultaneously sent into one single-mode cavity by Bob. Driven by a classical field, the atoms $A$, $D$ simultaneously interact with the single-mode cavity. Then, the atoms $B$, $E$ are simultaneously sent into the other single-mode cavity by Bob. Driven by a classical field, the atoms $B$, $E$ interact with the other single-mode cavity. Under the rotating-wave approximation, the interaction Hamiltonian between the single-mode cavity and the atoms can be expressed as[23-25]

$$H = \omega_0 S_z + \omega_a a^\dagger a + \sum_{j=1}^{2}[g(a^\dagger S_j^- + a S_j^\dagger) + \Omega(S_j^\dagger e^{-iwt} + S_j^- e^{iwt})], \quad (2)$$

where $S_z = (1/2)\sum_{j=1}^{2}(|e_j\rangle\langle e_j| - |g_j\rangle\langle g_j|)$, $S_j^- = |g_j\rangle\langle e_j|$, $S_j^\dagger = |e_j\rangle\langle g_j|$, $|g_j\rangle$ and $|e_j\rangle$ are the ground and excited states of the j$^{th}$ atom, $g$ is the atom-cavity coupling strength, $a$ and $a^\dagger$ are the annihilation and creation operators for the cavity mode, $\omega_0$, $\omega_a$, $\omega$, $\Omega$ are the atomic transition frequency, the cavity frequency, the classical field frequency and the Rabi frequency, respectively. Assume that $\omega_0 = \omega$. Accordingly, the evolution operator of the system in the interaction picture can be expressed as[23-25]

$$U(t) = e^{-iH_0 t} e^{-iH_e t}. \quad (3)$$

where $H_0 = \Omega\sum_{j=1}^{2}(S_j^\dagger + S_j^-)$, $H_e$ is the effective Hamiltonian. Considering the large detuning case $\delta \gg g$ ($\delta$ is the detuning between $\omega_0$ and $\omega_a$) and the strong driving regime $\Omega \gg \delta, g$, there is no energy exchange between the atomic system and the cavity. As a result, the effects of cavity decay and thermal field are eliminated. Then, in the interaction picture, the effective interaction Hamiltonian $H_e$ can be described by [23-25]

$$H_e = (\lambda/2)\left[\sum_{j=1}^{2}(|e_j\rangle\langle e_j| + |g_j\rangle\langle g_j|) + \sum_{i,j=1, i\neq j}^{2}(S_i^\dagger S_j^- + S_i^\dagger S_j^\dagger + H.C.)\right]. \quad (4)$$

where $\lambda = g^2/2\delta$. Rabi frequency and the interaction time are consistently chosen by Bob to satisfy $\Omega t = \pi$ and $\lambda t = \pi/4$ in both evolution cases. After the Hadamard operation ($H = (1/\sqrt{2})[(|g\rangle + |e\rangle)\langle g| + (|g\rangle - |e\rangle)\langle e|]$) is performed on the atom $C$, the total system will finally evolve into

$$|S^{-}\rangle_{ABC} \otimes |\psi^{-}\rangle_{DE} = (\sqrt{2}/4)[(-i|gg\rangle_{AD}|gg\rangle_{BE} + i|eg\rangle_{AD}|eg\rangle_{BE} + i|ge\rangle_{AD}|ge\rangle_{BE} - i|ee\rangle_{AD}|ee\rangle_{BE})|g\rangle_C + (-|gg\rangle_{AD}|ee\rangle_{BE} - |eg\rangle_{AD}|ge\rangle_{BE} + |ge\rangle_{AD}|eg\rangle_{BE} + |ee\rangle_{AD}|gg\rangle_{BE})|e\rangle_C]. \quad (5)$$

Extending $|\psi^-\rangle_{DE}$ to other three Bell states $|\psi^+\rangle_{DE}$, $|\phi^-\rangle_{DE}$ and $|\phi^+\rangle_{DE}$, if the evolving condition and process are the same as above, the total system will respectively evolve into

$$|S^-\rangle_{ABC} \otimes |\psi^+\rangle_{DE} = (\sqrt{2}/4)[(-|ee\rangle_{AD}|gg\rangle_{BE} - |eg\rangle_{AD}|ge\rangle_{BE}$$
$$+|ge\rangle_{AD}|eg\rangle_{BE} + |gg\rangle_{AD}|ee\rangle_{BE})|g\rangle_C + (i|gg\rangle_{AD}|gg\rangle_{BE} + i|eg\rangle_{AD}$$
$$|eg\rangle_{BE} + i|ge\rangle_{AD}|ge\rangle_{BE} + i|ee\rangle_{AD}|ee\rangle_{BE})|e\rangle_C], \qquad (6)$$

$$|S^-\rangle_{ABC} \otimes |\phi^-\rangle_{DE} = (\sqrt{2}/4)[(-i|gg\rangle_{AD}|ge\rangle_{BE} + i|eg\rangle_{AD}|ee\rangle_{BE}$$
$$+i|ge\rangle_{AD}|gg\rangle_{BE} - i|ee\rangle_{AD}|eg\rangle_{BE})|g\rangle_C + (-|gg\rangle_{AD}|eg\rangle_{BE} - |eg\rangle_{AD}$$
$$|gg\rangle_{BE} + |ge\rangle_{AD}|ee\rangle_{BE} + |ee\rangle_{AD}|ge\rangle_{BE})|e\rangle_C], \qquad (7)$$

$$|S^-\rangle_{ABC} \otimes |\phi^+\rangle_{DE} = (\sqrt{2}/4)[(-|eg\rangle_{AD}|gg\rangle_{BE} + |ge\rangle_{AD}|ee\rangle_{BE}$$
$$-|ee\rangle_{AD}|ge\rangle_{BE} + |gg\rangle_{AD}|eg\rangle_{BE})|g\rangle_C + (i|ge\rangle_{AD}|gg\rangle_{BE} + i|eg\rangle_{AD}$$
$$|ee\rangle_{BE} + i|ee\rangle_{AD}|eg\rangle_{BE} + i|gg\rangle_{AD}|ge\rangle_{BE})|e\rangle_C]. \qquad (8)$$

According to formulas (5)-(8), after the evolution of the total system, each result of the atoms $A$, $D$, the atoms $B$, $E$ and the atom $C$ do only correspond to one initial state among the above four known initial states. Four collections composed by different results of the atoms $A$, $D$, the atoms $B$, $E$ and the atom $C$ after evolution can be coded as:

$$\{|gg\rangle_{AD}|gg\rangle_{BE}|g\rangle_C, |eg\rangle_{AD}|eg\rangle_{BE}|g\rangle_C, |ge\rangle_{AD}|ge\rangle_{BE}|g\rangle_C,$$
$$|ee\rangle_{AD}|ee\rangle_{BE}|g\rangle_C, |gg\rangle_{AD}|ee\rangle_{BE}|e\rangle_C, |eg\rangle_{AD}|ge\rangle_{BE}|e\rangle_C,$$
$$|ge\rangle_{AD}|eg\rangle_{BE}|e\rangle_C, |ee\rangle_{AD}|gg\rangle_{BE}|e\rangle_C\} \to 00, \qquad (9)$$

$$\{|ee\rangle_{AD}|gg\rangle_{BE}|g\rangle_C, |eg\rangle_{AD}|ge\rangle_{BE}|g\rangle_C, |ge\rangle_{AD}|eg\rangle_{BE}|g\rangle_C,$$
$$|gg\rangle_{AD}|ee\rangle_{BE}|g\rangle_C, |gg\rangle_{AD}|gg\rangle_{BE}|e\rangle_C, |eg\rangle_{AD}|eg\rangle_{BE}|e\rangle_C,$$
$$|ge\rangle_{AD}|ge\rangle_{BE}|e\rangle_C, |ee\rangle_{AD}|ee\rangle_{BE}|e\rangle_C\} \to 11, \qquad (10)$$

$$\{|gg\rangle_{AD}|ge\rangle_{BE}|g\rangle_C, |eg\rangle_{AD}|ee\rangle_{BE}|g\rangle_C, |ge\rangle_{AD}|gg\rangle_{BE}|g\rangle_C,$$
$$|ee\rangle_{AD}|eg\rangle_{BE}|g\rangle_C, |gg\rangle_{AD}|eg\rangle_{BE}|e\rangle_C, |eg\rangle_{AD}|gg\rangle_{BE}|e\rangle_C,$$
$$|ge\rangle_{AD}|ee\rangle_{BE}|e\rangle_C, |ee\rangle_{AD}|ge\rangle_{BE}|e\rangle_C\} \to 01, \qquad (11)$$

$$\{|eg\rangle_{AD}|gg\rangle_{BE}|g\rangle_C, |ge\rangle_{AD}|ee\rangle_{BE}|g\rangle_C, |ee\rangle_{AD}|ge\rangle_{BE}|g\rangle_C,$$
$$|gg\rangle_{AD}|eg\rangle_{BE}|g\rangle_C, |ge\rangle_{AD}|gg\rangle_{BE}|e\rangle_C, |eg\rangle_{AD}|ee\rangle_{BE}|e\rangle_C,$$
$$|ee\rangle_{AD}|eg\rangle_{BE}|e\rangle_C, |gg\rangle_{AD}|ge\rangle_{BE}|e\rangle_C\} \to 10. \qquad (12)$$

Further extending $|S^-\rangle_{ABC}$ to other three GHZ states $|S^+\rangle_{ABC}$, $|P^+\rangle_{ABC}$ and $|P^-\rangle_{ABC}$, all the result collections composed by different results of the atoms $A$, $D$, the atoms $B$, $E$ and the atom $C$ after evolution from different initial states of the atoms $A$, $B$, $C$ and the atoms $D$, $E$ are listed in Table 1(1). Take the initial state of the atoms $A$, $B$, $C$ to be $|P^+\rangle_{ABC}^{10}$ and the initial state of the atoms $D$, $E$ to be $|\phi^+\rangle_{DE}^{10}$ for example. The superscripts in both $|P^+\rangle_{ABC}^{10}$ and $|\phi^+\rangle_{DE}^{10}$ denote that performing $U_2$ on the first atom can transform $|S^-\rangle_{ABC}$ to $|P^+\rangle_{ABC}$ and $|\psi^-\rangle_{DE}$ to $|\phi^+\rangle_{DE}$ respectively, while 11 denotes that the result collection composed by the result of the atoms $A$, $D$, the atoms $B$, $E$ and the atom $C$ after evolution from $|P^+\rangle_{ABC}$ and $|\phi^+\rangle_{DE}$ corresponds to formula (10).

In addition, If the atoms $A$, $B$, $C$ are in one of the four GHZ states including $|Q^\pm\rangle_{ABC} = (1/\sqrt{2})(|gge\rangle \pm |eeg\rangle)$ and $|R^\pm\rangle_{ABC} = (1/\sqrt{2})(|geg\rangle \pm |ege\rangle)$, and the atoms $D$, $E$ are still in one of the four Bell states, all the result collections composed by different results of the atoms $A$, $D$, the atoms $B$, $E$ and the atom $C$ after evolution from different initial states of the atoms $A$, $B$, $C$ and the atoms $D$, $E$ will be shown as Table 1(2).

Table 1. The result collections composed by different results of the atoms $A$, $D$, the atoms $B$, $E$ and the atom $C$ after evolution from different initial states of the atoms $A$, $B$, $C$ and the atoms $D$, $E$ (The superscript denotes the codes of $U_j$, $j = 0, 1, 2, 3$)

| (1) | | | | |
|---|---|---|---|---|
|  | $\|\psi^+\rangle_{DE}^{11}$ | $\|\psi^-\rangle_{DE}^{00}$ | $\|\phi^+\rangle_{DE}^{10}$ | $\|\phi^-\rangle_{DE}^{01}$ |
| $\|S^+\rangle_{ABC}^{11}$ | 00 | 11 | 01 | 10 |
| $\|S^-\rangle_{ABC}^{00}$ | 11 | 00 | 10 | 01 |
| $\|P^+\rangle_{ABC}^{10}$ | 10 | 01 | 11 | 00 |
| $\|P^-\rangle_{ABC}^{01}$ | 01 | 10 | 00 | 11 |

| (2) | | | | |
|---|---|---|---|---|
|  | $\|\psi^+\rangle_{DE}^{11}$ | $\|\psi^-\rangle_{DE}^{00}$ | $\|\phi^+\rangle_{DE}^{10}$ | $\|\phi^-\rangle_{DE}^{01}$ |
| $\|Q^+\rangle_{ABC}^{11}$ | 10 | 01 | 11 | 00 |
| $\|Q^-\rangle_{ABC}^{00}$ | 01 | 10 | 00 | 11 |
| $\|R^+\rangle_{ABC}^{10}$ | 00 | 11 | 01 | 10 |
| $\|R^-\rangle_{ABC}^{01}$ | 11 | 00 | 10 | 01 |

## 3 Quantum Steganography Protocol

The goal of our quantum steganography protocol is to covertly transmit secret messages from Alice to Bob by building up a hidden channel within normal quantum channel. The secret messages are grouped by three bits. Our quantum steganography protocol is illustrated in detail as follows.

S1) Preparation mode: A large number ($n$) of $|\psi^-\rangle_{DE}$ is prepared by Bob as the initial states of the atoms $D$, $E$. If the first bit from each group of secret messages is 1, a large number ($n$) of $|S^-\rangle_{ABC}$ will be prepared by Bob as the initial states of the atoms $A$, $B$, $C$, otherwise, a large number ($n$) of $|Q^-\rangle_{ABC}$ will be prepared by Bob as the initial states of the atoms $A$, $B$, $C$. Let $G_A$, $G_B$, $G_C$, $G_D$, $G_E$ represent the atom groups of $A$, $B$, $C$, $D$, $E$, respectively. Hence, $G_A = [A_1, A_2, \cdots, A_n]$, $G_B = [B_1, B_2, \cdots, B_n]$, $G_C = [C_1, C_2, \cdots, C_n]$, $G_D = [D_1, D_2, \cdots, D_n]$, $G_E = [E_1, E_2, \cdots, E_n]$. Bob transmits the atom $D$ in $G_D$ to Alice through quantum channel at first, and they perform eavesdropping checking together. Afterward, Bob transmits the atom $A$ in $G_A$ to Alice through quantum channel, and they perform eavesdropping checking together. Consequently, if there is no eavesdropping, Alice holds $G_A$, $G_D$, and Bob holds $G_B$, $G_C$, $G_E$.

S2) Control mode: (a) Eavesdropping checking for the transmission of the atom $D$ in $G_D$: 1) Alice selects a large enough subgroup from $G_D$, randomly measures the atom $D_i$ under the Z-basis $\{|g\rangle, |e\rangle\}$ or X-basis $\{|+\rangle, |-\rangle\}$, where $|+\rangle = \frac{1}{\sqrt{2}}(|g\rangle + |e\rangle)$ and $|-\rangle = \frac{1}{\sqrt{2}}(|g\rangle - |e\rangle)$, and publishes her measurement basis and corresponding measurement result to Bob through classical channel; 2) Bob measures $E_i$ from the corresponding subgroup of $G_E$ under the same basis; 3) By comparing with Alice's measurement result, Bob can know whether there is an eavesdropping or not. If the channel is safe, their measurement results are highly correlated. For example, when Alice and Bob perform Z-basis, Bob's measurement result should be $|g\rangle$ ($|e\rangle$) if Alice gets the measurement result $|e\rangle$ ($|g\rangle$). Then, if Bob confirms that there is an eavesdropping, they abort the communication; otherwise, they enter step (b); (b) Eavesdropping checking for the transmission of the atom $A$ in $G_A$: 1) Alice selects a large enough subgroup from $G_A$, randomly measures the atom $A_i$ under the Z-basis $\{|g\rangle, |e\rangle\}$ or X-basis $\{|+\rangle, |-\rangle\}$, and publishes her measurement basis and corresponding measurement result to Bob through classical channel; 2) Bob measures $B_i$ from the corresponding subgroup of $G_B$ and $C_i$ from the corresponding subgroup of $G_C$ under the same basis; 3) By comparing with Alice's measurement result, Bob can know whether there is an eavesdropping or not. If the channel is safe, their measurement results are highly correlated. Then, if Bob confirms that there is an eavesdropping, they abort the communication; otherwise, they enter information transmission mode.

S3) Information transmission mode: (a) According to the bits sequence of information, the unitary operations are performed by Alice on the atoms in $G_A$ and $G_D$, respectively (Note that after performed unitary operations, $G_A$ and $G_D$ turns to be $G_A'$ and $G_D'$, respectively. Although no unitary operations have been performed on the atoms in $G_B$, $G_C$ and $G_E$, for consistence, $G_B'$, $G_C'$, $G_E'$ are still used to represent the original $G_B$, $G_C$, $G_E$, respectively. Accordingly, $G_B'$ is the same as $G_B$, $G_C'$ is the same as $G_C$, and $G_E'$ is the same as $G_E$.); (b) Two atoms $A_m'$, $A_{m+1}'$ are chosen by Alice from $G_A'$ according to the behind two bits from each group of secret messages, and secret messages hiding mode follows; (c) $G_A'$ and $G_D'$ are sent back to Bob by Alice through quantum channel; (d) After $G_A'$ and the value of $m$ are obtained, GHZ-basis measurement is performed by Bob on $A_m'B_m'C_m'$ to recover the information it carried and the first bit from each group of secret messages.

S4) Secret messages hiding mode: (a) According to the behind two bits from each group of secret messages, two atoms $A_m'$, $A_{m+1}'$ are chosen by Alice from $G_A'$, where the subscript $m$ represents the position of the atom $A_m'$ in $G_A'$. The value of $m$ must satisfy the consistent condition, which means the GHZ initial state formed by $A_m'B_m'C_m'$ and the Bell initial state formed by $D_m'E_m'$ must be consistent with the behind two bits from each group of secret messages, as shown in Table 1(An appropriate $m$ can be decided by Alice in advance before sending $m$ to Bob by implementing QSDC or one-time pad through classical channel[21-22].); (b) By performing the same unitary operation on $A_{m+1}'$ in advance, $A_{m+1}'B_{m+1}'C_{m+1}'$ can copy the information carried by $A_m'B_m'C_m'$. Consequently, $A_{m+1}'B_{m+1}'C_{m+1}'$ doesn't normally transmit information but acts as an auxiliary GHZ state to help hide secret messages.

S5) The atoms $A_{m+1}'$, $D_m'$ are simultaneously sent into one single-mode cavity by Bob. Driven by a classical field, the atoms $A_{m+1}'$, $D_m'$ simultaneously interact with the single-mode cavity. Rabi frequency and the interaction time are chosen by Bob to satisfy $\Omega t = \pi$ and $\lambda t = \pi/4$.

S6) The atoms $B_{m+1}'$, $E_m'$ are simultaneously sent into the other single-mode cavity by Bob. Driven by a classical field, the atoms $B_{m+1}'$, $E_m'$ simultaneously interact with the other single-mode cavity. Rabi frequency and the interaction time are chosen by Bob to satisfy $\Omega t = \pi$ and $\lambda t = \pi/4$.

S7) Secret messages decoding mode: (a) The Hadamard operation is performed on the atom $C_{m+1}'$ by Bob; (b) After flying out the single-mode cavity, the states of the atoms $A_{m+1}'$, $D_m'$ are detected by Bob under the Z-basis $\{|g\rangle, |e\rangle\}$, respectively; (c) After flying out the other single-mode cavity, the states of the atoms $B_{m+1}'$, $E_m'$ are detected by Bob under the Z-basis $\{|g\rangle, |e\rangle\}$, respectively; (d) The state of the atom $C_{m+1}'$ is detected by Bob under the Z-basis $\{|g\rangle, |e\rangle\}$; (e) Based on the detection results of the atoms $A_{m+1}'$, $D_m'$, the atoms $B_{m+1}'$, $E_m'$, and the atom $C_{m+1}'$, the behind two bits from each group of secret messages can be decoded by Bob, according to formulas (9)-(12). Then, according to the state of $A_m'B_m'C_m'$, the first bit from each group of secret messages can be decoded by Bob. Finally, through the behind two bits from each group of secret messages together with the state of $A_m'B_m'C_m'$, the information carried by $D_m'E_m'$ can also be recovered by Bob, according to Table 1.

We use an example to explain our quantum steganography protocol. Assuming that a group of secret messages Alice wants to send Bob is **111**, and the information sequence …**11**$_{00}$…**00**$_{11}$…**10**$_{10}$…**01**$_{01}$…is generated by Alice (The information is divided by four bits, since two $U_j$ denote four bits information). In S1, a large number ($n$) of $|\psi^-\rangle_{DE}$ is prepared as the initial states of the atoms $D$, $E$, and a large number ($n$) of $|S^-\rangle_{ABC}$ is prepared as the initial states of the atoms $A$, $B$, $C$, since the first bit from this group of secret messages is **1**. In S3, the unitary operations are performed by

Alice on $G_A$ and $G_D$, respectively, according to the information sequence. Assume that the group numbers of $11_{00}$, $00_{11}$, $10_{10}$, $01_{01}$ in the information sequence are No.3, 7, 11 and 16, respectively. In S4, Alice can make $m = 3, 7, 11$ or 16 to satisfy the consistency in Table 1(1). If $m = 3$, $A'_3 B'_3 C'_3$ will be $|S^+\rangle$ and $D'_3 E'_3$ will be $|\psi^-\rangle$. Accordingly, the group of secret messages **111** is transmitted after entanglement swapping between $A'_4 B'_4 C'_4$ and $D'_3 E'_3$ in cavity QED together with Hadamard operation. The group of secret messages **111** can also be transmitted in the same way, if $m = 7, 11$ or 16. $A'_4 B'_4 C'_4$ can't be used to transmit the information like other normal GHZ states, and acts as an auxiliary GHZ state to help hide secret messages. In S7, the Hadamard operation is performed by Bob on the atom $C'_4$. Afterward, the states of the atoms $A'_4$, $D'_3$, the states of the atoms $B'_4$, $E'_3$ and the state of the atom $C'_4$ are detected by Bob under the Z-basis $\{|e\rangle, |g\rangle\}$, respectively. According to formulas (9)-(12), it can be decoded by Bob that the behind two bits from this group of secret messages are **11**. Then, according to the state of $A'_3 B'_3 C'_3$, it can be decoded by Bob that the first bit from this group of secret messages is **1**. Finally, through the behind two bits from this group of secret messages together with the state of $A'_3 B'_3 C'_3$, according to Table 1(1), it can be easily judged out by Bob that the information carried by $D'_3 E'_3$ is **00**.

## 4 Analysis
### 4.1 Capacity and Quantum Resource Used

In the above quantum steganography protocol, each three bits secret messages is transmitted by entanglement swapping between one GHZ state of $A'_{m+1} B'_{m+1} C'_{m+1}$ and one Bell state of $D'_m E'_m$ in cavity QED together with the Hadamard operation ($A'_{m+1} B'_{m+1} C'_{m+1}$ copies the information carried by $A'_m B'_m C'_m$ and acts as an auxiliary GHZ state to help hide secret messages). It is apparent that each three bits secret messages can be transmitted by four different kinds of initial states. For example, according to Table 1(1), **111** can be transmitted by four different kinds of initial states $\{|S^+\rangle_{ABC}^{11}, |\psi^-\rangle_{DE}^{00}\}$, $\{|S^-\rangle_{ABC}^{00}, |\psi^+\rangle_{DE}^{11}\}$, $\{|P^+\rangle_{ABC}^{10}, |\phi^+\rangle_{DE}^{10}\}$ and $\{|P^-\rangle_{ABC}^{01}, |\phi^-\rangle_{DE}^{01}\}$. After coding these four different kinds of initial states by formulas (13), the capacity of hidden channel in the above quantum steganography protocol can be increased to five bits. Consequently, its capacity of hidden channel is five times that of Ref.[17], Ref.[18], Ref.[19] and Ref.[20], respectively, and 1.25 times that of Ref.[21]. However, the capacity of Ref.[22], which achieves eight bits, is 1.6 times that of our quantum steganography protocol. It can be concluded that our quantum steganography protocol takes advantages over most of those previous quantum steganography protocols in the capacity of hidden channel.

$$1100 \rightarrow 00, 0011 \rightarrow 01, 1010 \rightarrow 10, 0101 \rightarrow 11. \quad (13)$$

Since all the quantum steganography protocols in Ref.[21], Ref.[22] and our protocol use entanglement swapping to build up a hidden channel, we make a comparison in the quantum resource used among them. Without considering the preparation mode and the control mode, in order to transmit five bits secret messages and four bits information successfully, our protocol needs to communicate three qubits (i.e., $A'_m$, $D'_m$ and $A'_{m+1}$) from Alice to Bob, and use five qubits (i.e., a GHZ state of $A'_m B'_m C'_m$ and a Bell state of $D'_m E'_m$) as normal quantum resource together with consuming three qubits (i.e., a GHZ state of $A'_{m+1} B'_{m+1} C'_{m+1}$) as auxiliary quantum resource in total. Moreover, five Z-basis measurements (i.e., measurements on $A'_{m+1}$, $D'_m$, $B'_{m+1}$, $E'_m$ and $C'_{m+1}$) and a GHZ-basis measurement (i.e., measurements on $A'_m B'_m C'_m$) are performed in total. In Ref.[21], without considering the preparation mode and the control mode, in order to transmit four bits secret messages and four bits information successfully, it needs to communicate three qubits (i.e., $A'_{m-1}$, $A'_m$ and $A'_{m+1}$) from Alice to Bob, and use four qubits (i.e., a Bell state of $A'_{m-1} B'_{m-1}$ and a Bell state of $A'_m B'_m$) as normal quantum resource together with consuming two qubits (i.e., a Bell state of $A'_{m+1} B'_{m+1}$) as an auxiliary quantum resource in total. Moreover, three Bell-basis measurements (i.e., measurements on $A'_m A'_{m+1}$, $B'_m B'_{m+1}$ and $A'_{m-1} B'_{m-1}$) are performed in total. In Ref.[22], without considering the preparation mode and the control mode, in order to transmit eight bits secret messages and eight bits information successfully, it needs to communicate six qubits (i.e., $P_1^{m-1}$, $P_3^{m-1}$, $P_1^m$, $P_3^m$, $P_1^{m+1}$ and $P_3^{m+1}$) from Alice to Bob, and use eight qubits (i.e., a $\chi$-type state of $|\chi^{ij}\rangle_{3214}^{m-1}$ and a $\chi$-type state of $|\chi^{pq}\rangle_{3214}^{m}$) as normal quantum resource together with consuming four qubits (i.e., a $\chi$-type state of $|\chi^{ij}\rangle_{3214}^{m+1}$) as an auxiliary quantum resource in total. Moreover, two CMB-basis measurements (i.e., measurements on particle groups $[P_3^m, P_1^{m+1}, P_2^m, P_4^{m+1}]$ and $[P_3^{m+1}, P_1^m, P_2^{m+1}, P_4^m]$) and one FMB-basis measurement (i.e., measurement on $|\chi^{ij}\rangle_{3214}^{m-1}$) are performed in total. It can be concluded that the capacity of Ref.[22] is larger than that of our protocol and that of Ref.[21] at the cost of more quantum resource used.

### 4.2 Imperceptibility

In our quantum steganography protocol, the choice of $m$ is not arbitrary for Alice, since the value of $m$ must satisfy the consistent condition with respect to $A'_m B'_m C'_m$, $D'_m E'_m$ and the behind two bits from each group of secret messages. Consequently, the imperceptibility mainly depends on the difficulty of knowing $m$ by Eve. As pointed out in Ref.[21-22], choosing $m$ can still be treated as arbitrary behave for Eve, since both information and secret messages can be regarded to be random or pseudo-random. If information or secret messages doesn't distribute randomly in advance, pseudo-random sequence encryption can be adopted to make its distribution randomized.

For example, if a group of secret messages Alice wants to send Bob is **111**, in order to choose $m$, Alice needs to find out all the group numbers of "**11**$_{00}$","**00**$_{11}$","**10**$_{10}$","**01**$_{01}$" in the information sequence. Accordingly, the states of $A_m'B_m'C_m'$ and $D_m'E_m'$ will be "$|S^+\rangle|\psi^-\rangle$", "$|S^-\rangle|\psi^+\rangle$", "$|P^+\rangle|\phi^+\rangle$" and "$|P^-\rangle|\phi^-\rangle$", respectively. If the information distributes evenly, probability of "**11**$_{00}$","**00**$_{11}$","**10**$_{10}$", "**01**$_{01}$" will be $1/16$, respectively. Consequently, their total probability is $1/4$. It is straightforward if a group of secret messages is **100**,**101** or **110**. Therefore, the probability distributions of information and secret messages will make $m$'s uncertainty best, according to Shannon's information theory, as pointed out in Ref.[21-22]. Consequently, choosing $m$ can be regarded to be random for Eve. It means that the imperceptibility of our quantum steganography protocol is good.

### 4.3 Security

The security of our quantum steganography protocol is based on the security for the transmission of $G_D$ and $G_A$ from Bob to Alice.

Now we analyze the security for the transmission of the atom $D$ in $G_D$ from Bob to Alice at first. According to Stinespring dilation theorem, the eavesdropping of Eve can be realized by a unitary operation $\hat{E}$ on a larger Hilbert space, $|x,E\rangle \equiv |x\rangle|E\rangle$. Then, the state of the composite system composed by Bob, Alice and Eve is

$$|\psi\rangle = \sum_{a,b\in\{g,e\}} |\varepsilon\rangle|a\rangle|b\rangle, \quad (14)$$

where $|\varepsilon\rangle$ denotes Eve's probe state, $|a\rangle$ and $|b\rangle$ are single-atom states of Alice and Bob in each Bell state, respectively. The condition on the states of Eve's probe is

$$\sum_{a,b\in\{g,e\}} \langle\varepsilon|\varepsilon\rangle = 1. \quad (15)$$

Eve can only eavesdrop $G_D$ before the first checking, and Eve's effect on the system will be

$$\hat{E}|g,E\rangle = \hat{E}|g\rangle|E\rangle = \alpha_1|g\rangle|\varepsilon_{00}\rangle + \beta_1|e\rangle|\varepsilon_{01}\rangle, \quad (16)$$

$$\hat{E}|e,E\rangle = \hat{E}|e\rangle|E\rangle = \beta_1'|g\rangle|\varepsilon_{10}\rangle + \alpha_1'|e\rangle|\varepsilon_{11}\rangle. \quad (17)$$

Then, the whole system will be

$$|\psi\rangle = \frac{1}{\sqrt{2}}[(\alpha_1|g\rangle|\varepsilon_{00}\rangle + \beta_1|e\rangle|\varepsilon_{01}\rangle)|e\rangle - (\beta_1'|g\rangle|\varepsilon_{10}\rangle + \alpha_1'|e\rangle|\varepsilon_{11}\rangle)|g\rangle], \quad (18)$$

where $\varepsilon_{00}$, $\varepsilon_{01}$, $\varepsilon_{10}$, $\varepsilon_{11}$ are Eve's states, respectively. Moreover, Eve's probe operator can be expressed as

$$\hat{E} = \begin{pmatrix} \alpha_1 & \beta_1' \\ \beta_1 & \alpha_1' \end{pmatrix}. \quad (19)$$

Since $\hat{E}$ is a unitary operator, the complex numbers $\alpha_1$, $\beta_1$, $\alpha_1'$ and $\beta_1'$ should satisfy

$$E\hat{E}^\dagger = 1. \quad (20)$$

Consequently, we can obtain the following relations:

$$|\alpha_1|^2 = |\alpha_1'|^2, |\beta_1|^2 = |\beta_1'|^2. \quad (21)$$

Then, the error rate introduced by Eve's eavesdropping on $G_D$ will be

$$\tau_1 = |\beta_1|^2 = |\beta_1'|^2 = 1 - |\alpha_1|^2 = 1 - |\alpha_1'|^2. \quad (22)$$

Therefore, Eve's eavesdropping on $G_D$ will inevitably introduce an error rate, and be discovered by Alice and Bob.

Without loss of generality, we take the example of measurement-resend attack on $G_D$ to further explain the error rate introduced by Eve's eavesdropping. Eve intercepts the atom $D$ in $G_D$, randomly measures it in Z-basis or X-basis, and resends his measurement result to Alice. In the first case, Eve performs Z-basis measurement. The state of the whole system will collapse to $|g\rangle_D|e\rangle_E$ or $|e\rangle_D|g\rangle_E$ each with probability $1/2$. Take the state to be $|g\rangle_D|e\rangle_E$ for example. Accordingly, Eve resends $|g\rangle_D$ to Alice. Then, if Alice performs Z-basis measurement to check eavesdropping, no error will be introduced by Eve. If Alice performs X-basis measurement, the state collapses to $|+\rangle_D|+\rangle_E$, $|+\rangle_D|-\rangle_E$, $|-\rangle_D|+\rangle_E$ or $|-\rangle_D|-\rangle_E$ each with probability $1/4$. As a result, the error rate introduced by Eve will be $50\%$. Therefore, the total error rate in this case is $25\%$. In the second case, Eve performs X-basis measurement. The state of the whole system will collapse to $|+\rangle_D|-\rangle_E$ or $|-\rangle_D|+\rangle_E$ each with probability $1/2$. Take the state to be $|+\rangle_D|-\rangle_E$ for example. Accordingly, Eve resends $|+\rangle_D$ to Alice. Then, if Alice performs Z-basis measurement to check eavesdropping, the state collapses to $|g\rangle_D|g\rangle_E$, $|g\rangle_D|e\rangle_E$, $|e\rangle_D|g\rangle_E$ or $|e\rangle_D|e\rangle_E$ each with probability $1/4$. As a result, the error rate introduced by Eve will be $50\%$. If Alice performs X-basis measurement, no error will be introduced by Eve. Therefore, the total error rate in this case is $25\%$. Therefore, the random Z-basis or X-basis measurement guarantees that Eve's measurement-resend attack on $G_D$ can be found out by eavesdropping check.

The security for the transmission of $G_A$ from Bob to Alice can be analyzed in a similar way as above. After Eve eavesdrop $G_A$ before the second checking, the whole system will be

$$|S\rangle = \frac{1}{\sqrt{2}}[(\alpha_2|g\rangle|\varepsilon_{00}\rangle + \beta_2|e\rangle|\varepsilon_{01}\rangle)|ee\rangle - (\beta_2'|g\rangle|\varepsilon_{10}\rangle + \alpha_2'|e\rangle|\varepsilon_{11}\rangle)|gg\rangle] \quad (23)$$

or

$$|Q\rangle = \frac{1}{\sqrt{2}}[(\alpha_2|g\rangle|\varepsilon_{00}\rangle + \beta_2|e\rangle|\varepsilon_{01}\rangle)|ge\rangle - (\beta_2'|g\rangle|\varepsilon_{10}\rangle + \alpha_2'|e\rangle|\varepsilon_{11}\rangle)|eg\rangle]. \quad (24)$$

Finally, by deducing in the same way as above, we can know that the error rate introduced by Eve's eavesdropping on $G_A$ will be

$$\tau_2 = |\beta_2|^2 = |\beta_2'|^2 = 1 - |\alpha_2|^2 = 1 - |\alpha_2'|^2. \quad (25)$$

It also can be concluded that, Eve's eavesdropping on $G_A$ will inevitably introduce an error rate, and be discovered by Alice and Bob.

Without loss of generality, we take the example of measurement-resend attack on $G_A$ to further explain the error rate introduced by Eve's eavesdropping. Assume that $|S^-\rangle_{ABC}$ is prepared by Bob as the initial states of the atoms $A$, $B$, $C$. Eve intercepts the atom $A$ in $G_A$, randomly measures it in Z-basis or X-basis, and resends his measurement result to Alice. In the first case, Eve performs Z-basis measurement. The state of the whole system will collapse to $|g\rangle_A|e\rangle_B|e\rangle_C$ or $|e\rangle_A|g\rangle_B|g\rangle_C$ each with probability $1/2$. Take the state to be $|g\rangle_A|e\rangle_B|e\rangle_C$ for example. Accordingly, Eve resends $|g\rangle_A$ to Alice. Then, if Alice performs Z-basis measurement to check eavesdropping, no error will be introduced by Eve. If Alice performs X-basis measurement, the state collapses to $|+\rangle_A|+\rangle_B|+\rangle_C$, $|+\rangle_A|+\rangle_B|-\rangle_C$, $|+\rangle_A|-\rangle_B|+\rangle_C$, $|+\rangle_A|-\rangle_B|-\rangle_C$, $|-\rangle_A|+\rangle_B|+\rangle_C$, $|-\rangle_A|+\rangle_B|-\rangle_C$, $|-\rangle_A|-\rangle_B|+\rangle_C$ or $|-\rangle_A|-\rangle_B|-\rangle_C$ each with probability $1/8$. As a result, the error rate introduced by Eve will be 50%. Therefore, the total error rate in this case is 25%. In the second case, Eve performs X-basis. The state of the whole system will collapse to $|+\rangle_A|+\rangle_B|-\rangle_C$, $|+\rangle_A|-\rangle_B|+\rangle_C$, $|-\rangle_A|+\rangle_B|+\rangle_C$ or $|-\rangle_A|-\rangle_B|-\rangle_C$ each with probability $1/4$. Take the state to be $|+\rangle_A|+\rangle_B|-\rangle_C$ for example. Accordingly, Eve resends $|+\rangle_A$ to Alice. Then, if Alice performs Z-basis measurement to check eavesdropping, the state collapses to $|g\rangle_A|g\rangle_B|g\rangle_C$, $|g\rangle_A|g\rangle_B|e\rangle_C$, $|g\rangle_A|e\rangle_B|g\rangle_C$, $|g\rangle_A|e\rangle_B|e\rangle_C$, $|e\rangle_A|g\rangle_B|g\rangle_C$, $|e\rangle_A|g\rangle_B|e\rangle_C$, $|e\rangle_A|e\rangle_B|g\rangle_C$ or $|e\rangle_A|e\rangle_B|e\rangle_C$ each with probability $1/8$. As a result, the error rate introduced by Eve will be 75%. If Alice performs X-basis measurement, no error will be introduced by Eve. Therefore, the total error rate in this case is 37.5%. Therefore, the random Z-basis or X-basis measurement guarantees that Eve's measurement-resend attack on $G_A$ can be found out by eavesdropping check. If $|Q^-\rangle_{ABC}$ is prepared by Bob as the initial states of the atoms $A$, $B$, $C$, the same conclusion can be drawn with respect to Eve's measurement-resend attack on $G_A$.

Further consider the influence caused by leakage of $m$. Assume that Eve not only obtains $m$ but also gets $D'_m$ and $A'_{m+1}$ through some advance eavesdropping attacks (It is possible for Eve to eavesdrop $D'_m$ and $A'_{m+1}$, since the atom $D$ and the atom $A$ are travel atoms). However, Eve still can't get secret messages, because Eve doesn't know that there is Hadamard operation performed on the atom $C'_{m+1}$ by Bob. Moreover, even though Eve knows the Hadamard operation, according to formulas (9)-(12), only knowing $D'_m$ and $A'_{m+1}$ is not enough to decode secret messages.

## 5 Conclusion

To sum up, we present a large payload quantum steganography protocol based on cavity QED, which effectively uses the evolution law of atom in cavity QED. The protocol builds up hidden channel to transmit secret messages using entanglement swapping between one GHZ state and one Bell state in cavity QED together with the Hadamard operation. It is insensitive to cavity decay and thermal field. Its imperceptibility and security can be guaranteed. Furthermore, its capacity of hidden channel achieves five bits, larger than most of those previous quantum steganography protocols.